\documentclass{article}
\usepackage{ijcai24}

\usepackage{times}
\usepackage{soul}
\usepackage{url}
\usepackage[hidelinks]{hyperref}
\usepackage[utf8]{inputenc}
\usepackage[small]{caption}
\usepackage{graphicx}
\usepackage{amsmath}
\usepackage{amsthm}
\usepackage{amsmath,amssymb,amsfonts}
\usepackage{booktabs}
\usepackage[switch]{lineno}
\usepackage{bm}
\usepackage{threeparttable} 
\usepackage{wrapfig}
\usepackage{algorithm}
\usepackage{algorithmic}
\usepackage{subcaption}
\usepackage{autobreak}

\title{FedMPQ: Secure and Communication-Efficient Federated Learning with Multi-codebook Product Quantization}

\author{
Xu Yang$^1$
\and
Jiapeng Zhang$^1$
\and
Qifeng Zhang$^1$
\and
Zhuo Tang$^1$
\affiliations
$^1$Hunan University \\
\emails
\{yangxv, zhangjp,zqfhnucsee,ztang\}@hnu.edu.cn,
}

\begin{document}

\maketitle
\def\x{\mathbf x}
\def\y{\mathbf y}
\def\z{\mathbf z}
\def\h{\mathbf h}
\def\p{\mathbf p}

\def\hx{{\hat{\mathbf x}}}
\def\hy{{\hat{\mathbf y}}}
\def\hz{{\hat{\mathbf z}}}

\def\tz{{\tilde{\mathbf z}}}
\def\tc{{\tilde{\mathbf c}}}

\def\bbE{\mathbb E}
\def\bbR{\mathbb R}

\begin{abstract}
In federated learning, particularly in cross-device scenarios, secure aggregation has recently gained popularity as it effectively defends against inference attacks by malicious aggregators. However, secure aggregation often requires additional communication overhead and can impede the convergence rate of the global model, which is particularly challenging in wireless network environments with extremely limited bandwidth. Therefore, achieving efficient communication compression under the premise of secure aggregation presents a highly challenging and valuable problem.
In this work, we propose a novel uplink communication compression method for federated learning, named FedMPQ. 
Specifically, we utilize updates from the previous round to generate multiple robust codebooks. Secure aggregation is then achieved through trusted execution environments (TEE) or a trusted third party (TTP).
In contrast to previous works, our approach exhibits greater robustness especially in scenarios where the data is not independently and identically distributed (non-IID) and there is a lack of sufficient public data. 
The experiments conducted on the LEAF dataset demonstrate that our proposed method achieves 99\% of the uncompressed baseline's final accuracy, while reducing the uplink communications by 90-95\%.
\end{abstract}

\section{Introduction}


%
Federated Learning (FL) is a secure distributed learning paradigm designed to train a global model using data dispersed across a large-scale and distributed group of edge devices~\cite{kairouz2021advances}. 
Rather than simply transmitting raw data, federated learning involves an iterative process of parameter sharing between devices and a server through heterogeneous networks. Besides, the number of participating devices can be extensive, while being connected to the internet in a slow or unstable manner.


In the design of conventional cross-device federated learning systems, during each training round, the server initiates the process by broadcasting the model to the clients. 
Subsequently, clients conduct training based on their respective data and update the model using gradient descent algorithms. 
The updates are then transmitted back to the server, which aggregates them considering updates from all clients in the current round. In the context of cross-device federated learning, hundreds to thousands of personal/IoT devices actively participate in each training round. 
Consequently, given the characteristics of edge devices, several pressing challenges need immediate attention in cross-device federated learning. 

First and foremost, the efficiency of communication, especially in terms of the uploading process, becomes paramount in cross-device federated learning. 
Under the mainstream federated learning network architecture like a mobile network, the downlink bandwidth is typically $8-20$ times greater than the uplink bandwidth~\cite{lee2012mobile}, leading to a high congestion in the parameter uploading process.
Second, all clients are considered stateless, as individual clients participate in either a single round or several discontinuous rounds of the whole training process.
Third, these edge devices may have limited computational capabilities, which consequently become unbearable to the excessive computational overhead. 
Fourth, client data is usually non-IID and highly sensitive in terms of privacy~\cite{49232}.
\begin{figure}
    \centering
    \includegraphics[width=0.99\linewidth]{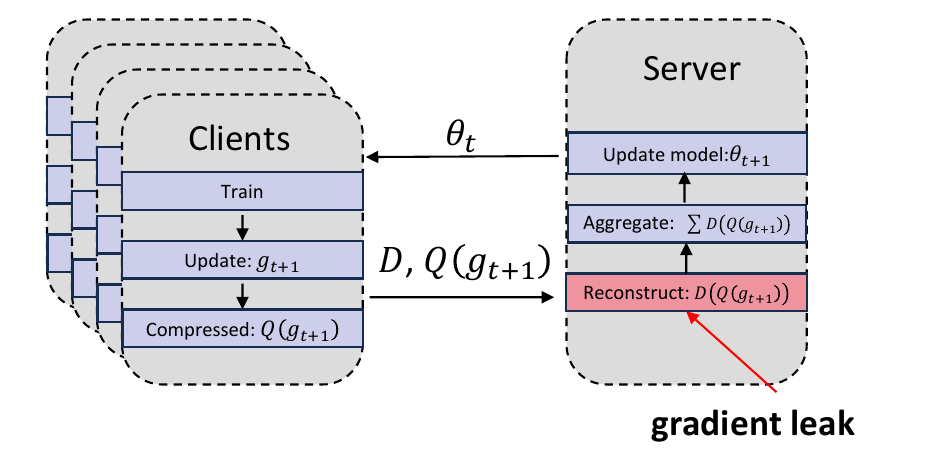}
    \caption{Gradient leak.
The gradient updates from clients are exposed to the server, allowing the server to potentially infer certain features of the training data by analyzing these updates. 
Where $D$ represents the client-customized reconstruction operation.}
    \label{fig:leak}
\end{figure}
In light of the aforementioned challenges, the following question has become a focal point in federated learning: 
\textbf{how to reduce uplink communications and preserve model accuracy, all under the umbrella of ensuring user privacy?}
Numerous ongoing efforts explore the application of gradient compression algorithms in federated learning to mitigate communication overhead. 
These efforts draw inspiration primarily from data-center-based training methods such as QSGD~\cite{alistarh2017qsgd}, Pruning~\cite{sahu2021rethinking,zhangfedduap} and product quantization~\cite{dai2019hyper}. 
While these algorithms demonstrate significant effectiveness and broad applicability, they often lack customized adaptations tailored to the distinctive characteristics of federated learning. This limitation not only impacts the convergence rate as well as the accuracy of the global model, but also introduces potential privacy vulnerabilities.
As shown in Figure~\ref{fig:leak}, all clients employ specialized compressors to compress the updates before uploading them. 
Consequently, the server must individually reconstruct each client's updates before the aggregation step, leading to a risk of gradient leakage. 
It is imperative to prevent this situation, as an honest yet inquisitive server might be able to infer the training data employed in the current round from the reconstructed gradient.

\label{conditions}To tackle this, we present a secure and communication-efficient federated learning method with multi-codebook product quantization.
In contrast to previous arts~\cite{oh2022communication,prasad2022reconciling} which use traditional product quantization or rely solely on public data, we combine local public data with the previous round's updates to predict several shared codebooks for the upcoming round. 
Specifically, during each training round, clients not only upload compressed updates but also contribute pseudo-codebooks generated from their updates.  The server then generates multiple shared codebooks based on these pseudo-codebooks, which are employed for the subsequent round of training. 

This approach empowers us to obtain more robust codebooks and mitigates the impact of non-IID public data. 
To safeguard data privacy, updates are aggregated within a TEE or by a TTP in the compressed domain before being decompressed, ensuring that no single party can reconstruct the updates. 
Furthermore, we introduce a pruning-based error residual compression strategy, allowing clients to flexibly control the compression rate.
Therefore, it achieves data security and uplink communication compression under the common but stringent real-world scenarios, such as: (\textbf{i}) Clients require flexible control over the uplink communication compression rate to adapt to varying network conditions.
(\textbf{ii}) The server maintains a small amount of public data, but it differs significantly from client data.
(\textbf{iii}) Participating training clients are stateless, have low device capabilities, and are numerous.
(\textbf{iv}) The compression method must be suitable for secure aggregation to prevent gradient leakage.


In summary, we present the key contributions in this paper:
\begin{itemize}

\item We introduce FedMPQ, a novel multi-codebook product quantization compression method, which combines local public data and client updates to generate multiple codebooks.
\item 
 To the best of our knowledge, we are the first to optimize product quantization with a residual error
 pruning process, addressing the issue of inflexibility in controlling the compression rate inherent in product quantization.
\item We also propose a novel aggregation scheme in which client updates are sent for aggregation in the TEE or by a TTP in the compressed domain. 
This approach ensuring that neither the server nor the third party can reconstruct the source data.

\item The experiments demonstrate that our proposed method achieves a higher convergence rate compared to previous works under a similar compression ratio without compromising the trained model’s accuracy. The method even works better in the scenarios with non-IID data.
\end{itemize}

\begin{figure*}
    \centering
      \vspace{-4mm}
      \includegraphics[width=0.95\textwidth]{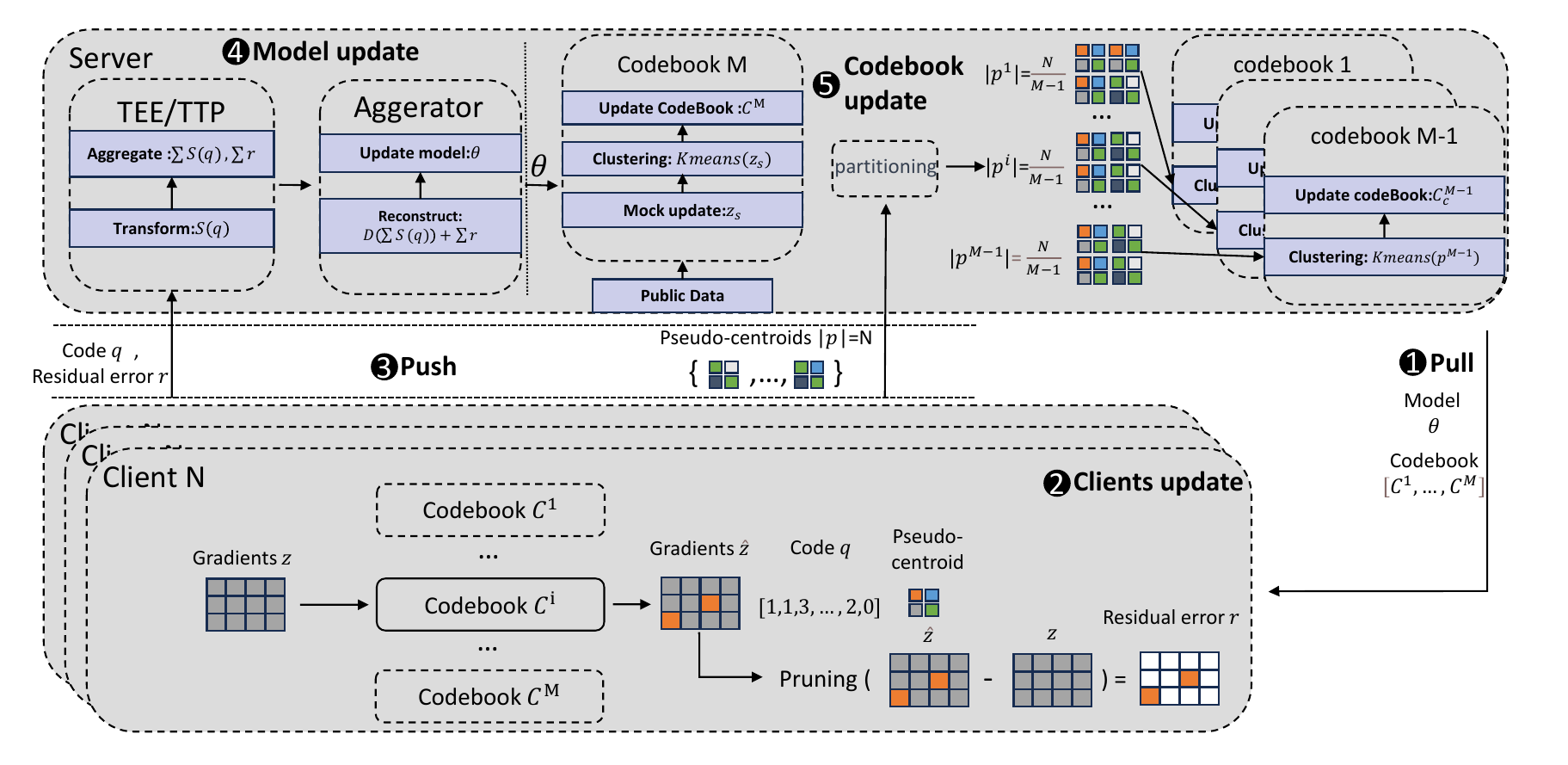}
    \caption{The training process of FedMPQ Framework. In each training round, $N$ clients participate, and each client uploads residual error, code, and pseudo-centroids. For further details, please refer to section~\ref{Proposed}. Best view in color.}
    \label{fig:framework}
\end{figure*}
\section{Preliminaries}
\subsection{Cross-device Federated Learning}
Federated learning is a machine learning setting where multiple clients collaborate in solving a machine learning problem, under the coordination of a central server. 
Further, cross-device FL represents a more challenging and prevalent scenario within the field. 
In addition to common constraints such as non-IID and privacy-sensitive data, this setting involves a large number of participating devices with limited computational capabilities and possibly unstable network conditions. 
Consequently, the balance among the privacy security, communication efficiency and model accuracy stands as a central objective in the realm of cross-device federated learning.

\subsection{Federated Learning with Server Learning}
Federated learning faces performance challenges when the training data among clients is non-IID. 
A novel complementary approach to mitigate this performance degradation involves leveraging a small public dataset when the server has access to it. The server can employ various techniques, such as auxiliary learning based on the public data~\cite{mai2022federated} or dynamic hyperparameter tuning~\cite{zhangfedduap}, to address this issue.
This approach has also been incorporated into communication compression strategies~\cite{prasad2022reconciling}, where the server simulates the next-round updates using a public dataset. 
These simulated updates are then used to tailor the next round's communication compression strategy. 
When the distribution of server data is similar to that of the aggregate samples of all clients, this strategy yields significant benefits. 
However, accurately determining the overall sample distribution of clients in practice remains a demanding task.
\subsection{Secure Aggregation}
In federated learning, it is possible to gain insights into the training data through methods such as membership inference or reconstruction attacks targeted at a trained model~\cite{carlini2022membership,9767718}. 
Additionally, the transmission of local updates can inadvertently lead to data exposure~\cite{geiping2020inverting}. 
To prevent the server from directly inspecting updates from each client, which is a primary avenue for information leakage, the Secure Aggregation protocol has been introduced.
Secure Aggregation is a widely used component in large-scale cross-device federated learning implementations, allowing server aggregation of client updates without requiring individual access to them. 
While it does not provide complete prevention against client data leakage, it remains a practical protocol in cross-device federated learning where user communication and computational resources are limited~\cite{kairouz2021advances}. 

\subsection{Communication Compress}
Communication compression is a crucial aspect of cross-device federated learning since clients often use unreliable and slow wireless networks to upload and receive parameters. 
Uplink communication poses more challenges compared to downlink communication~\cite{prasad2022reconciling}. 
PowerSGD~\cite{vogels2019powersgd} introduced an efficient low-rank gradient compressor, which improves convergence rate through error feedback. Methods such as FedVQCS~\cite{oh2023fedvqcs} employ PQ-based compression strategies, achieving high compression ratios. 
However, the aforementioned approaches are not compatible with the secure aggregation paradigm, limiting their applicability. 
While compression methods proposed in FetchSGD~\cite{pmlr-v119-rothchild20a} and \cite{prasad2022reconciling} can be adapted to secure aggregation protocols, there is still room for improvement in scenarios where client data is non-IID.
%

\section{Proposed Method}\label{Proposed}
In the context of federated learning, given an original uncompressed parameter set $G = {g_i}_{i=1}^N$ that includes updates from $N$ clients in a round, we aim to construct operations $(\mathcal{Q}, \mathcal{D})$, representing compression and decompression, respectively. 
The objective is to fulfill the following condition:
\begin{equation}
  \mathcal{D}(\frac{1}{N}\sum_{i=1}^{N}\mathcal{Q}(g_i)) = \frac{1}{N}\sum_{i=1}^{N}\mathcal{D}(\mathcal{Q}(g_i)) \approx \frac{1}{N} \sum_{i=1}^{N} g_i.    
\end{equation}
The equation implies that we can exchange the operation of aggregation and decompression without performance degradation, thereby avoiding the server obtaining the uncompressed updated parameters from any individual user. 

For this purpose, we propose a multi-codebook product quantization compression strategy named FedMPQ.
As shown in Figure~\ref{fig:framework}, the training process of FedMPQ consists of multiple rounds, each of which is composed of five steps:

\textbf{(1) Pull}: Clients download parameters, including the complete model parameters $\theta$ and $M$ different codebooks.

\textbf{(2) Client Update}: In this step, $N$ clients participating in the current round compute gradients z based on their local data. 
Each client attempts to compress these gradients using $M$ codebooks, selecting the one with the minimum error as the result. 
Finally, the clients output the compressed gradients, encoding, pseudo-centroids, and residual error.

\textbf{(3) Push}: Clients upload the quantization codes $q$, residuals $r$, and pseudo-centroids $p$ to the server.

\textbf{(4) Model Update}: The server aggregates the updates securely in the compressed domain using a secure aggregation protocol and then directly recovers the aggregated updates.

\textbf{(5) Codebook Update}: The server leverages public data to simulate the training process, generating one codebook. It also uniformly splits the pseudo-centroids uploaded by clients into $M-1$ parts and clusters them to update the other $M-1$ codebooks.
    
In the sequel, we elaborate the motivation and details of our proposed method, where the basic steps of Pull and Push are omitted.

\subsection{Client update}

In previous research~\cite{prasad2022reconciling}, public data were employed to approximate the update process on clients and generate shared codebooks. However, when there is a substantial disparity between the distribution of public data and client data, model accuracy may significantly deteriorate. Therefore, in FedMPQ, on the basis of generating codebooks using public data from the server, we aim to enable clients to upload partial information to assist the server in codebook generation without compromising privacy. 
The overall client-side algorithm is outlined in Algorithm~\ref{alg:client_compression}.

\begin{algorithm}[t]
\caption{Client Update}
\label{alg:client_compression}
\begin{algorithmic}[1]
\REQUIRE Global model parameters $\theta$ and $M$ codebooks.
\STATE Initialize variables: $\hat{z}_i, p_i, r_i$ for $i \in \{1, 2, \ldots, M\}$
\FOR {$i = 1$ to $M$}
    \STATE Train local model with local data and obtain update gradient $z$
    \STATE Compress the gradient using codebook $i$:\\ $q_i \gets \text{Compress}(z, \text{Codebook}_i)$
    \STATE Compute quantization error: \\ $r_i \gets z - \text{Decompress}(q_i, \text{Codebook}_i)$
    \STATE Calculate pseudo-centroid: \\ $p_i \gets \text{CalculatePseudoCentroid}(z, q_i, \text{Codebook}_i)$
\ENDFOR
\STATE Select the codebook with the smallest total residual: $i^* = \arg\min_i \sum_j |r_i[j]|$
\STATE Prune the top-k residuals and their positions: \\ $r_{i^*} \gets \text{Pruning}(r_{i^*}, k)$
\STATE Upload pruned residuals $r_{i^*}$, quantization code $q_{i^*}$, pseudo-centroid $p_{i^*}$, and $i^*$ to the server
\end{algorithmic}
\end{algorithm}
\subsubsection{Product Quantization}\label{sec:pq}
The essence of Product Quantization (PQ) is to decompose the high-dimensional vector space into the Cartesian product of subspaces and then quantize these subspaces separately \cite{5432202}. The product quantizer consists of $M$ codebooks 
$\mathcal{C} =[\bm{C}^1,\ldots, \bm{C}^M]$, 
where the $n$-th codebook 
$\bm{C}^n \in{R}^{K \times D} $ 
contains $K$ codewords 
$\bm{C}^n =
[\bm{c}^n_1,\ldots,\bm{c}^n_K]$ and each codeword is a $D$-dimensional cluster-centroid. 
For a parameter vector $\bm{z} \in {R}^{L}$ of a model layer, if we choose to compress it using the 
$n$-th codebook, the first step involves partitioning vector $\bm{z} \in {R}^{L}$ from the encoding stage into $\frac{L}{D}$ subvectors: $\bm{z} = [\bm{z}^1,\cdots,\bm{z}^m,\cdots,\bm{z}^\frac{L}{D}]$. In the $m$-th subspace, the product quantization approximates $\bm{z}^m$ by the nearest codeword:
\begin{gather}
  \hat{\bm{z}}^m_{n} =\arg \min_{q^m\in[K]}||\bm{z}^m_n - \bm{c}^n_{q^m}||.
\end{gather}
Then, we obtain quantization code and the quantized reconstruction feature $\hat{\bm{z}}$ as:
\begin{equation}
  {q}_n=\left[ {q}^1_n,{q}^2_n,\cdots,{q}^{\frac{L}{D}}_{n}\right],
  \label{eq:index}
\end{equation}
\begin{equation}
  \hat{\bm{z}}_{n} =\left[ \hat{\bm{z}}^1_{n},\hat{\bm{z}}^2_{n},\cdots,\hat{\bm{z}}^\frac{L}{D}_{n} \right],
\end{equation}
where ${q}$ can be represented as a binary code with $M\log_2(K)$ bits. 
In FedMPQ, the parameter vector $\bm{z}$ attempts to be quantized using all codebooks, and the one with the smallest distance to $\bm{z}$ is selected as the final result  $\hat{\bm{z}}$:
\begin{gather}
  \hat{\bm{z}} =\arg \min_{n\in[M]}||\bm{z} - \hat{\bm{z}}_n||.
\end{gather}
Furthermore, it is essential to consider that neural networks exhibit significant variations in the parameters across different layers. 
Therefore, we maintain a separate set of codebooks for each layer of the neural network to minimize errors.
\subsubsection{pseudo-centroid generated}
Generating a sufficiently robust shared codebook solely from the server's public data is challenging because we cannot accurately simulate the data distribution of the client-side private data. 
To address this issue, during the Push process in each training round, we aim to upload certain parameters that assist the server in generating the codebook for the next round while ensuring that gradients are not leaked. 
Hence, we introduce a simple and efficient approach that enables the simultaneous generation of the requisite codewords during the vector quantization of updates in federated learning.
%

Assume that there are $n_i^m$ subvectors approximated by codeword $\bm{c}_i^m$, which are denoted as $\{\bm{s}_1, \cdots, \bm{s}_{n_i^m}\}$, where $m$ denotes the $m$-th codebook of the quantizer and $i$ denotes the $i$-th codeword of the codebook. We then update the codeword $\bm{c}_i^m$ to approach the mean value of these subvectors:
\begin{align}
\begin{split}
  \bm{c}_i^m &\gets \bm{c}_{i}^m \cdot(1-\gamma) + \frac{\sum_j^{n_i^m} \bm{s}_j \cdot \gamma}{n_i^m},\label{eq:ema2}\\  %
\end{split}
\end{align}    
where $\gamma$ is a discount factor corresponding to the learning rate, which is initialized as 0.99. Subsequently, the clients will select the top half of codewords, chosen the most times, as pseudo-centroids to upload to the server.

\subsubsection{Residual Error Pruning}
PQ has two drawbacks. 
First, it suffers from bias in compression, inevitably leading to errors. 
Previous methods have often mitigated this issue by maintaining an error accumulation vector at the client, but this solution is challenging to apply in cross-device federated learning since clients are stateless. Second, PQ's compression rate can hardly be flexibly controlled, making it challenging to dynamically adjust data transmission based on the actual client communication bandwidth. 
To address these two issues, we propose a supplementary residual compression scheme based pruning. 
By controlling the magnitude of uploaded residuals, we can simultaneously reduce errors and flexibly adjust the communication volume, allowing for better adaptation to complex real-world scenarios. 
Experimental results demonstrate that residuals ranging from $0.1\%$ to $5\%$ significantly enhance model convergence rates.
%
\subsection{Cloud Update}
\subsubsection{Secure Aggregation}
In regards to the problem of compressing client-to-server model updates under the Secure Aggregation primitive, \cite{prasad2022reconciling} has provided a detailed discussion and proposed a viable solution. 
Building upon their SECIND approach, we have made some enhancements. 
Algorithm~\ref{alg:qaa} illustrates the entire process of secure aggregation, with all sensitive operations conducted within TEE or TPP. 
The aggregator can only access $O_{final}$ and $R_{final}$, thus effectively preventing gradient leakage. 

\begin{algorithm}[t]
\caption{Aggregation Algorithm}
\label{alg:qaa}
\begin{algorithmic}[1]

\REQUIRE $N$ clients with quantization codes, pruned residuals, and shared codebook $C$.
\ENSURE Final aggregated $g$
\STATE // \textit{TEE/TPP Phase: Aggregation in compress domain} 
\STATE Initialize an empty matrix $O_{\text{final}}, R_{\text{final}}$.
\FOR{$i = 1$ to $N$}
\STATE Initialize an empty matrix $O_i$.

\FOR{each element $e$ in client $i$'s quantization code}
    \STATE Perform one-hot encoding on $e$ to create a row vector $v_e$ in $\mathbb{R}^{1 \times K}$.
    \STATE Concatenate $v_e$ row-wise to matrix $O_i$.
\ENDFOR
\STATE Reconstruct pruned residuals for client $i$ to generate $R_i$

\STATE $R_{\text{final}} \gets R_{\text{final}}+R_i$
\STATE $O_{\text{final}} \gets O_{\text{final}}+O_i$.
\ENDFOR

\STATE // \textit{Aggregator Phase: Reconstruct
}

\STATE Reconstruct aggregated gradient $g \gets C \times O_{\text{final}} + R_{\text{final}}$.

\RETURN $g$.

\end{algorithmic}
\end{algorithm}


\subsubsection{Codebook update}
Applying PQ to gradient compression poses a challenge since all PQ algorithms necessitate learning the codebook for compressing gradients. 
Nevertheless, we cannot acquire the gradients essential for codebook learning until the commencement of model training.
An ideal solution is to simulate the client training process using public data to obtain gradients. 
However, this approach is not applicable in scenarios where data is non-IID, as there may be significant differences between the public dataset and the clients' private data. 
To tackle this problem, we aim to estimate the next-round gradients by analyzing and utilizing the model's historical information during training, especially the gradient information from the previous round. 
To achieve this, we maintain multiple codebooks. One of these codebooks is computed using publicly available data, while the remaining codebooks are calculated utilizing pseudo-centroids uploaded by the clients. 
Algorithm~\ref{alg::federated_codebook} presents the complete flow of the algorithm.

\subsection{Performance Discussion}
\subsubsection{Convergence Discussion}
The compression technique introduced by FedMPQ exhibits a bias. 
For biased compression operations, represented as $Q(\cdot)$,~\cite{li2023analysis} have shown that biased compressed SGD converges when $Q$ is a $\tau$-contraction: 
\begin{equation}
  \label{eq:convergence}
  ||Q(x) - x|| \leq \tau ||x|| , 0 \leq \tau < 1,
\end{equation}
It is evident that PQ satisfies this condition, as long as it is enforced that there exists a zero vector within the codebook. 
Furthermore, empirical experiments demonstrate that our proposed approach exhibits no significant degradation in both accuracy and convergence rate when compared to the uncompressed baseline.
\subsubsection{Communication and Computation Discussion}
In our proposed FedMPQ, clients download the complete global model and all shared codebooks at the beginning of each round of training. 
After local training is completed, they upload quantization codes, residual error and pseudo-centroids equivalent to half the size of a single codebook. 
The codebooks are relatively small compared to the original gradients, making the additional overhead negligible. 
Consequently, our method significantly reduces the uplink communication cost, which is cost-effective in wireless network environments. 
Regarding computational expenses, the most time-consuming distance calculations in quantization can be optimized using matrix multiplication, resulting in an additional computational cost of $\mathcal{O}(MLK)$ for a given vector to be quantized $z \in R^L$, where $M,K$ has the same meaning as in Section \ref{sec:pq}. 
This cost is acceptable compared to the computational overhead of client model training.
\begin{algorithm}[h]
\caption{Federated Codebook Generation}
\label{alg::federated_codebook}
\begin{algorithmic}[1]
\REQUIRE Pseudo-centroids $p_1, \ldots, p_N$, codebook count $n$, codebook size $K$, codeword length $D$, public dataset
\ENSURE $M$ codebooks
\STATE Initialize an empty list $codebooks$
\STATE Simulate training process on public data to obtain a model gradient
\STATE Apply k-means algorithm to generate $K$ code vectors of length $D$ for the first codebook using the simulated gradient
\STATE Append the first codebook to $codebooks$
\STATE Divide the $N$ Pseudo-centroids into $M-1$ equal parts
\FOR{$i=1$ to $M-1$}
    \STATE Sample $n$ Pseudo-centroids from the $i$-th part
    \STATE Apply k-means algorithm to generate $K$ code vectors of length $D$ for the $i$-th codebook using the sampled Pseudo-centroids
    \STATE Append the $i$-th codebook to $codebooks$
\ENDFOR
\RETURN $codebooks$
\end{algorithmic}
\end{algorithm}

\section{Evaluation}
In this section, we introduce the datasets, parameter choices and models we experiment on. 
Then we thoroughly analyse the performance of the proposed method with comprehensive comparisons and ablation studies.

\subsection{Experimental Setup}
\subsubsection{Datasets and Metrics}
As mentioned in Section~\ref{conditions}, our experiments predominantly concentrated on tackling communication compression challenges prevalent in typical yet challenging scenarios encountered in cross-device federated learning:
(\textbf{i}) Non-IID client data.
(\textbf{ii}) A substantial number of clients with relatively modest individual data volumes.
(\textbf{iii}) A restricted quantity of data retained by the server, characterized by significant disparities compared to client data.


To ensure that our experiments adhere to the aforementioned conditions, we assess the effectiveness of our approach using the following two datasets sourced from the LEAF benchmark~\cite{caldas2018leaf}: CelebA~\cite{liu2015deep} and Femnist~\cite{deng2012mnist}. 
For the CelebA dataset, which comprises 2 classes, the data was partitioned into 9343 clients in a non-IID manner, with each client having an average of 21 samples. We randomly selected 20 images as public data. For the Femnist dataset, encompassing 62 classes, the data was partitioned into 1775 clients, with each client having an average of 88 samples. Furthermore, a collection of 60 images was selected to serve as public data.

We comprehensively evaluate FedMPQ from three perspectives: convergence rounds, peak model accuracy, and total communication cost with respect to the target accuracy. 
%
Since downlink bandwidth is usually $8-20$ times greater than the uplink bandwidth~\cite{lee2012mobile}, without loss of generality, we calculate the total communication cost with
\begin{equation}
  \label{eq:comm_cost}
   cost_{total} = \frac{cost_{downlink}}{8} + cost_{uplink}.
\end{equation}
\subsection{Implementation Details}
We implement FedMPQ based on FLSim and conducted experiments comparing three methods: pruning, scalar quantization (SQ), and single-codebook product quantization (SPQ)~\cite{prasad2022reconciling}. We conducted multiple experiments for each item to reduce errors.

For the CelebA dataset, we conducted training using a convolutional binary classifier~\cite{nguyen2022federated}, with a model size of 114KB. Each training round randomly selected 100 clients to participate. The local learning rate of $\eta_{l}$ was set to 0.9, while the server learning rate $\eta$ was set to 0.08. 
Regarding the Femnist dataset, we opted for ResNet-18 as the backbone for the classification of 62 categories, resulting in a model size of 43MB. In each training round, 30 clients were chosen to participate, with a local learning rate of $\eta_{l}$ set to 0.01, and a server learning rate of $\eta$ set to 0.24.

Due to the limited amount of client data, multiple training rounds can easily lead to overfitting.
To avoid this issue, each client undergoes a single full training epoch in every training round.

\subsection{Experimental Results}
\begin{figure}
  \centering
  \begin{subfigure}{0.495\linewidth}
    \includegraphics[width=1\linewidth]{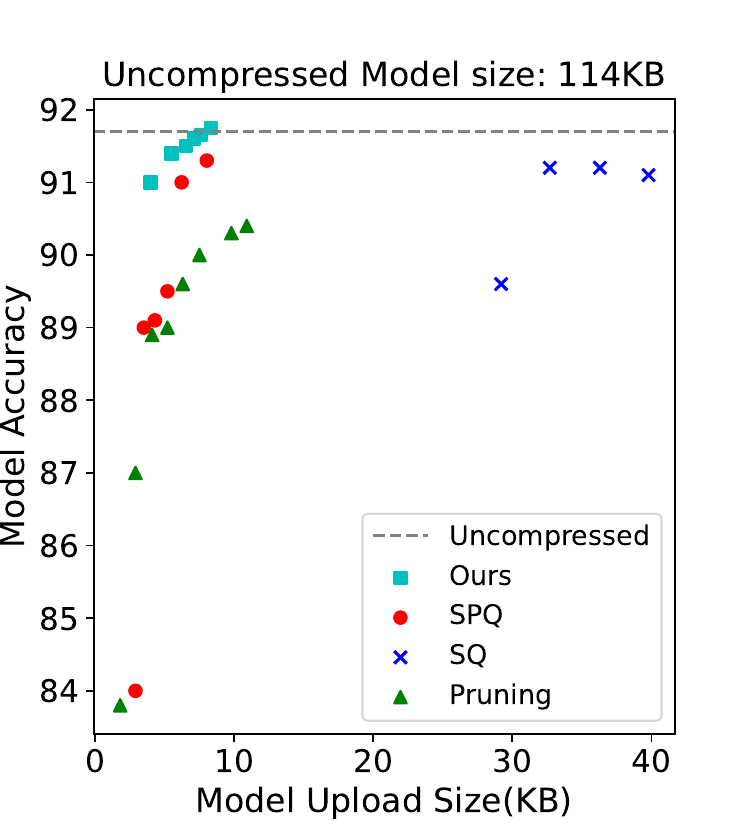}
    \caption{CelebA}
    \label{fig:celeba_scatter}
  \end{subfigure}
  \begin{subfigure}{0.495\linewidth}
    \includegraphics[width=1\linewidth]{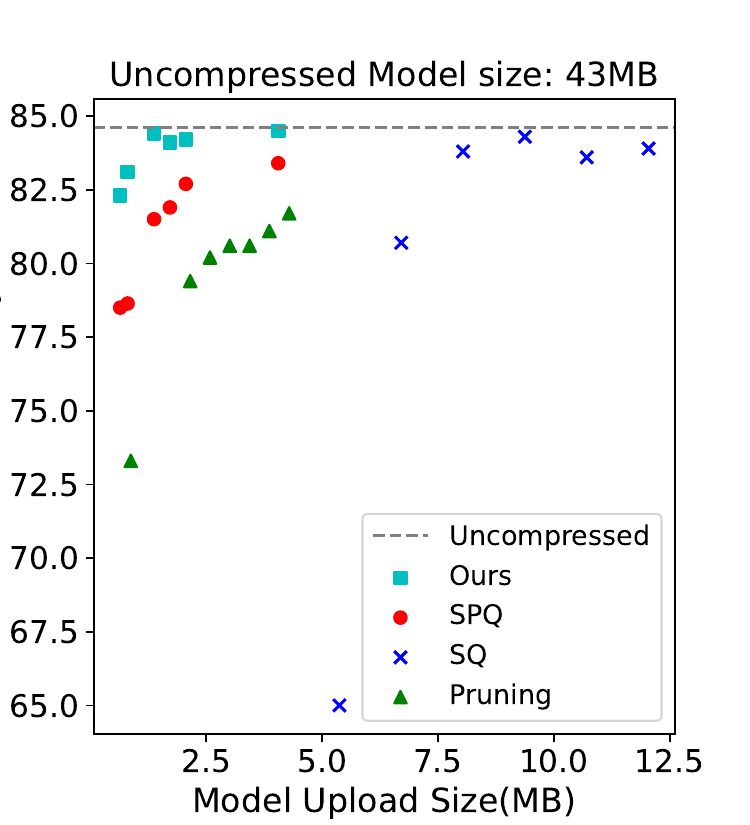}
    \caption{Femnist}
    \label{fig:enter-label2}
  \end{subfigure}
  \caption{The relationship between model accuracy and communications, with both FedMPQ and PQ employing the same limited public dataset for codebook training, and both uploading 0.1\% of residual error. ``Uncompressed" denotes the results obtained without any compression. }
  \label{fig:main_result}
\end{figure}

\subsubsection{Uplink Communications vs. Model Accuracy} 
The primary motivation behind proposing FedMPQ is to minimize uplink communication. Consequently, we have provided the model accuracy and uplink communication curves in Figure~\ref{fig:main_result}. We have conducted experiments for each compression strategy at various compression ratios. 
Each strategy is trained until the global model converges, and the highest model accuracy is taken as the final result. 
From the results, the following observations can be made:
\textbf{(i)}
 Our proposed FedMPQ demonstrates the best performance in terms of uplink communication compression and final model accuracy curves. On both the CelebA and Femnist datasets, FedMPQ achieves approximately 25x uplink communication compression with minimal loss in accuracy. Under extreme compression ratios, all methods exhibit a significant decline in model accuracy, but FedMPQ still outperforms others noticeably.
\textbf{(ii)} In comparison to FedMPQ, SPQ exhibits a noticeable decline in model accuracy, indicating that codebooks generated solely from limited public data cannot effectively compress client updates.


\subsubsection{Total Communication Cost vs. Model Accuracy}
\begin{figure}
  \centering
  \begin{subfigure}{0.495\linewidth}
    \includegraphics[width=1\linewidth]{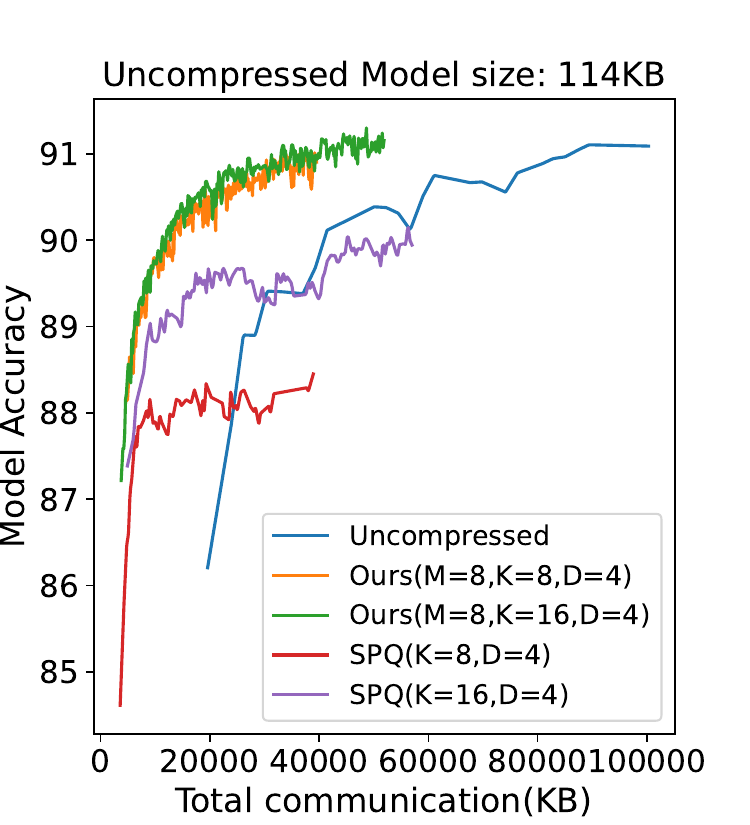}
    \caption{CelebA}
    \label{fig:celeba_scatter}
  \end{subfigure}
  \begin{subfigure}{0.495\linewidth}
    \includegraphics[width=1\linewidth]{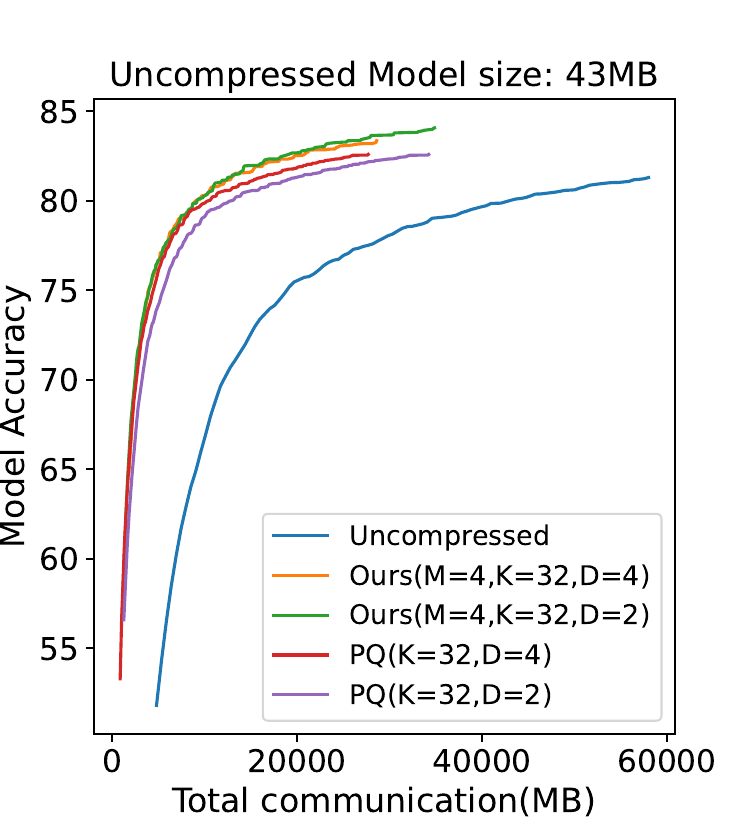}
    \caption{Femnist}
    \label{fig:feminst_scatter}
  \end{subfigure}
    \caption{Model Accuracy vs. Total communication cost. $M$ is the number of codebooks, $K$ represents the number of codewords in each codebook, and $D$ stands for the length of each codeword. }
    \label{fig:enter-label}
\end{figure}
In federated learning, a higher uplink communication compression rate can impair the convergence rate, leading to an increase in the number of training rounds required to achieve the desired accuracy. Consequently, we have presented curves depicting the relationship between total communication cost and model accuracy. Since the performance of pruning and SQ is significantly lower than that of SPQ and FedMPQ, we only show the results for SPQ and FedMPQ. Figure~\ref{fig:celeba_scatter} depicts the results on the CelebA dataset, and it is evident from the figure that FedMPQ outperforms SPQ significantly. SPQ exhibits a significant drop in model accuracy compared to the baseline without compression, and it is highly unstable. This indicates that compressing the model only on a limited amount of public data can severely compromise the effectiveness of federated learning when training the generator model on a small scale. Figure~\ref{fig:feminst_scatter} illustrates the performance of our method on the Femnist dataset. Although FedMPQ and SPQ can both demonstrate noticeable effects, FedMPQ still achieves superior performance to SPQ. This indicates that, even on large-scale models, our method can still accelerate the convergence rate and yield better final model accuracy compared to SPQ.

\subsubsection{Ablation study}

\begin{figure}
  \centering
  \begin{subfigure}{0.495\linewidth}
  \includegraphics[width=1\linewidth]{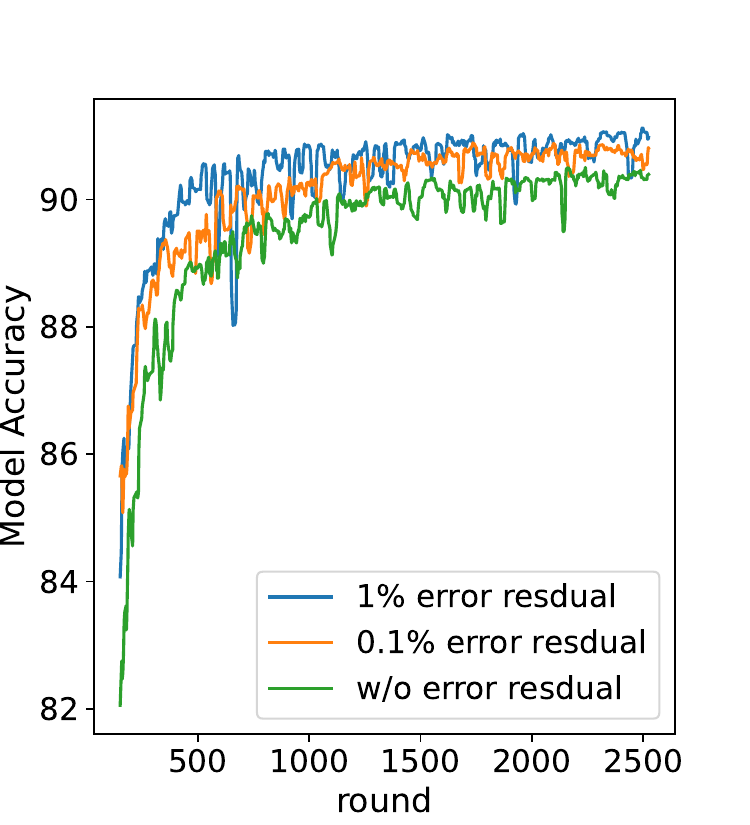}
    \caption{CelebA}
    \label{fig:celeba_ef}
  \end{subfigure}
  \begin{subfigure}{0.495\linewidth}
    \includegraphics[width=1\linewidth]{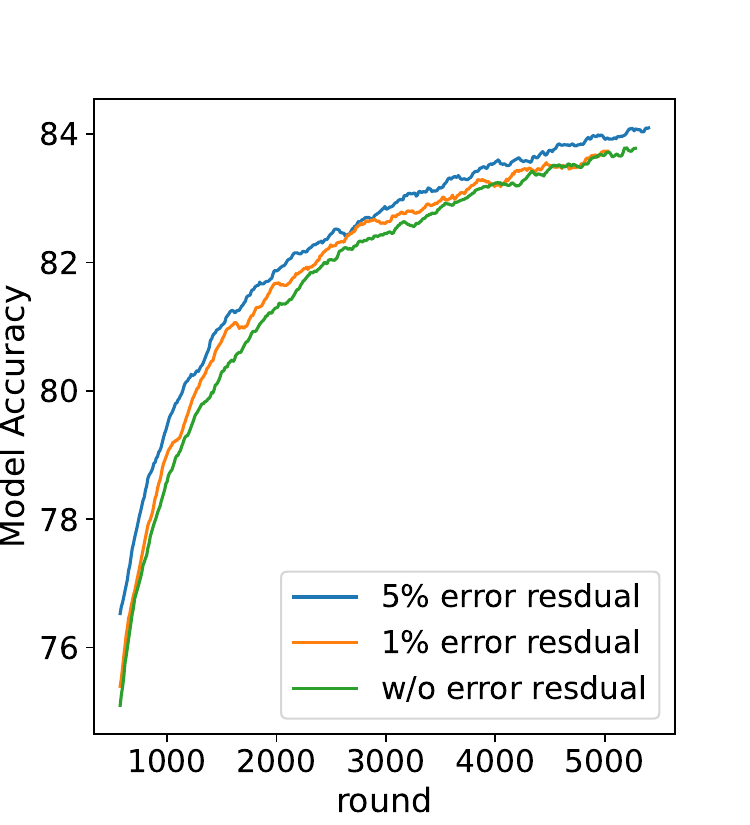}
    \caption{Femnist}
    \label{fig:minist_ef}
  \end{subfigure}
  \caption{Effects of residual error upload ratio. On the CelebA dataset, we set $M=8, K=8, D=4$. On the Femnist dataset, $M=4, K=32, D=4$. Where $M,K,D$ has the same meaning as in Section \ref{sec:pq}.}
\label{fig:enter-label}
\end{figure}

We investigate the impact of residual error compression rates on FedMPQ performance. To facilitate quantitative analysis, we fix all clients to upload the same ratio of residuals. 
As shown in Figure~\ref{fig:celeba_ef}, On the CelebA dataset, uploading only 0.1\% of the residual error effectively enhances the convergence rate and final model accuracy. 
Figure~\ref{fig:minist_ef} presents the results on the Femnist dataset, where uploading 1\% of the residual error slightly improves the convergence rate. 
When uploading 5\% of the residual error, both convergence rate and final model accuracy exhibit observable improvements. 
Hence, different codebook shapes and models can influence the contribution of residual error to the overall performance of FedMPQ.
\begin{figure}
    \centering
    \vspace{-5mm}
    \includegraphics[width=0.90\linewidth]{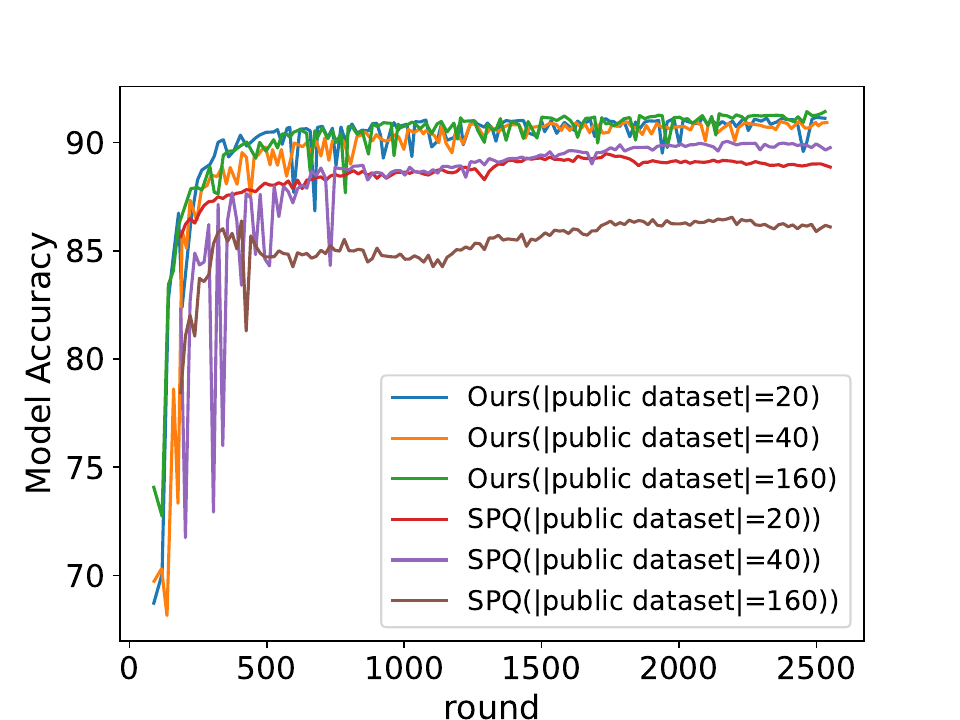}
    \caption{Effects of server dataset size. $M=8, K=8, D=4$.}
    \label{fig:dataset}
\end{figure}


Figure~\ref{fig:dataset} illustrates the impact of the number of public data on training effectiveness on the CelebA dataset. We randomly selected 20, 40, and 160 images as the public dataset and displayed the convergence curves for both PQ and FedMPQ. Across all three dataset sizes, FedMPQ consistently exhibited stable performance. However, for SPQ, the performance across all three datasets is relatively poor, particularly when the dataset size reaches 160. This observation suggests that indiscriminate expansion of the public dataset's size does not necessarily enhance codebook quality. As mentioned in~\cite{mai2022federated}, the distribution of the public data needs to closely align with the client data to achieve better results. However, this alignment is often challenging to achieve due to the non-IID of client data.
\begin{table}
    \centering
    \begin{tabular}{cccccc}
        \toprule
        M  & K & D& Data num & Rounds to 90\% Acc \\
        \midrule
4 & 32 & 2&20 &  $288\pm20$ \\
8 & 16 & 2&20 &  $338\pm18$ \\
8 & 8 & 2&20 &  $426\pm70$ \\
8 & 32 & 4&20 & $427\pm85$ \\
8 & 16& 4&20 &  $547\pm81$ \\
8 & 8&  4&20 &  $711\pm40$ \\
4 & 8&  8&20 & $766\pm80$ \\
8 & 4&  4&20 &  $819\pm90$ \\
7 & 8&  4&w/o public data& $2900\pm130$ \\
1 & 8&  4&20 & $\infty$ \\
\midrule
\multicolumn{4}{c}{Uncompressed}&  $225\pm 35$ \\
        \bottomrule
    \end{tabular}
    \caption{Effects of codebook shape ($K,D$) and number ($M$) on CelebA. ``uncompressed" refers to the results obtained without applying any communication compression methods, ``w/o public data" indicates that no public data was used, and ``$M=1$" signifies the absence of pseudo-centroids. }
    \label{tab:table1}
\end{table}

Table~\ref{tab:table1} presents the impact of different codebook shapes and quantities on the final model accuracy and convergence rate. From the experimental results, it is evident that both codebook quantity and shape significantly affect the model's ultimate performance and convergence rate.
When the codebooks are sufficiently large, FedMPQ is capable of achieving performance similar to the uncompressed baseline while exhibiting very similar convergence rates. 
Furthermore, ``M=7" denotes results obtained exclusively from pseudo-centroids, while ``M=1" reflects those derived solely from public data. While using only pseudo-centroids to generate codebooks still allows the model to achieve a 90\% accuracy rate, there is a significant decrease in convergence rate. This leads to a substantial increase in total communication overhead, underscoring the indispensability of codebooks generated based on public data.
\section{Conclusion}
In this paper, we have addressed the challenge of realizing secure and efficient communication compression in federated learning scenarios where there exists a significant gap between the public and client data. We thus introduced FedMPQ, an innovative approach that goes beyond previous product quantization-based compression methods. In FedMPQ, we maintain and update multiple codebooks simultaneously, allowing clients to adaptively select a best codebook to compress their updates. Recognizing the inflexibility of traditional product quantization in adjusting compression rates, we complemented our approach with a pruning-based residual compression method, enabling clients to finely tune compression rates according to their available bandwidth. Besides, the separation of model and codebook update ensures a secure aggregation. 
Our experiments have demonstrated that these strategies can be effectively integrated to achieve substantial uplink compression, minimal performance degradation, and enhanced security guarantees.

\bibliographystyle{named}
\bibliography{ijcai24}

\end{document}